\documentclass{aa}
\usepackage{graphicx,aabib}
\def \ob{CXOGC\,J174540.0-290031}

\newcommand{\oversim}[2]{\lower0.5ex\vbox{\baselineskip=0pt\lineskip=0.2ex
     \ialign{$\mathsurround=0pt #1\hfil##\hfil$\crcr#2\crcr\sim\crcr}}}
\newcommand{\simless} {\mbox{$\mathrel{\mathpalette\oversim<}$}} 
\begin{document}

\title{Discovery of X-ray eclipses from the transient source \ob{}
 with XMM-Newton}

\author{D. Porquet\inst{1} 
           \and N. Grosso\inst{2}
            \and G. B\'elanger\inst{3}
            \and A. Goldwurm\inst{3}
             \and F. Yusef-Zadeh\inst{4}
            \and R.S. Warwick\inst{5}
            \and P. Predehl\inst{1} 
       }
\offprints{Delphine Porquet\\ (dporquet@mpe.mpg.de)}

\institute{Max-Planck-Institut f\"{u}r extraterrestrische Physik,
P.O. Box 1312, Garching bei M\"{u}nchen D-85741, Germany
\and Laboratoire d'Astrophysique de Grenoble, Universit\'e Joseph-Fourier,
 BP53, 38041 Grenoble Cedex 9, France
\and CEA Saclay, DSM/DAPNIA/SAp, 91191 Gif-sur-Yvette Cedex, France
\and Department of Physics and Astronomy, Northwestern University, Evanston, IL 60208, USA 
\and Department of Physics and Astronomy, University of Leicester, Leicester LE1 7RH, UK
}
\date{Received ... ; Accepted ...  }

\abstract{We present the {\sl XMM-Newton} observations obtained during 
four revolutions  in Spring and Summer 2004  of \ob,
 a moderately bright transient X-ray source, 
 located at only 2.9$^{\prime\prime}$ from Sgr\,A*. We report the discovery of
  sharp and deep X-ray eclipses, with a period of 27,961$\pm$5\,s and a duration
of about 1,100$\pm$100\,s, observed during the two consecutive {\sl XMM-Newton} 
 revolutions from August 31 to September 2. 
 No deep eclipses were present during the two consecutive {\sl XMM-Newton} 
 revolutions  from March 28 to April 1, 2004.
 The spectra during all four observations 
are well described with an absorbed  power law continuum.
 While our fits on the power law index  over the four observations
 yield values that are consistent with $\Gamma$=1.6--2.0, there appears to
be a significant increase in the column density during the Summer 2004
observations, i.e. the period during which the eclipses are detected.
  The intrinsic luminosity in the 2--10\,keV energy range is almost constant
 with 1.8--2.3 $\times$ 10$^{34}\,({\rm d_{8\,kpc}})^{2}$\,erg\,s$^{-1}$
  over the four observations.  
In the framework of eclipsing semidetached binary systems, we show that the
  eclipse period constrains the mass of the assumed main-sequence
 secondary star to less than 1.0\,M$_\odot$.
 Therefore, we deduce that \ob{} is a low-mass X-ray binary (LMXB).
 Moreover the eclipse duration constrains the mass of the compact object
 to less than about 60\,M$_\odot$, which is consistent
 with a stellar mass black hole or a neutron star. 
 The absence of deep X-ray eclipses during the Spring 2004 observations 
could be explained if the centroid of the X-ray
emitting region moves from a position on the orbital plane
to a point above the compact object, possibly coincident with
the base of the jet which was detected in radio at this epoch.
According to our study, \ob{} is a LMXB, and is more likely to 
have a black hole than a neutron star as its primary,
which would entail an inclination angle greater than 75$^{\circ}$, 
i.e. the binary system and the accretion disk are seen close to edge-on.
}
\titlerunning{Discovery of X-ray eclipses from \ob{}}
\authorrunning{Porquet et al.}
\maketitle

\section{Introduction}

The Galactic center region is a very complex region of the 
Galaxy as viewed in X-ray wavelengths.
 This region contains numerous diffuse and 
point-like sources, fluorescent X-ray emission from molecular 
clouds,  supernova remnants, compact stellar sources as well as  young 
and evolved stellar clusters.  In addition to Sgr A*, the compact radio 
source which is thought to coincide with a  supermassive black hole
at the dynamical center of the Galaxy (\cite{Ghez03}, \cite{Sch03}), 
this region is populated by accreting compact objects, such as neutron
stars and black hole binaries which are often transient in nature
with peak X-ray luminosities in excess of 10$^{36}$\,erg\,s$^{-1}$ 
(e.g., \cite{Ch97}, \cite{Si99}, \cite{Sa02}, \cite{P03a}).
Repeated, high sensitivity X-ray observations of this region carried
out by {\sl XMM-Newton} and {\sl Chandra}, primarily for the purpose of
monitoring Sgr A*, have also lead to the discovery of several
 X-ray transient binaries in various accretion states
 from quiescence (L$_{\rm X} <$ 10$^{33}$\,erg\,s$^{-1}$) to   
 bright outburst (L$_{\rm X} >$ 10$^{36}$\,erg\,s$^{-1}$) as reported 
by \cite*{Sa05}, \cite*{P05}, and \cite*{muno05a}. 
Transients with intermediate X-ray luminosities
 ($\sim$10$^{34}$erg\,s$^{-1}$)  
appear to be over-abundant by a factor of about 20
 per unit stellar mass within 1\,pc of Sgr\,A* (\cite{muno05a}). 

During the large {\sl XMM-Newton} multi-wavelength project to monitor
the light curve of Sgr A*, observations were carried out
 in the Spring and Summer of 2004,
 a brightening in the 2--10\,keV energy band of
a factor of about two was detected within a radius of 
10$^{\prime\prime}$ around SgrA* (\cite{B05}).
The centroid of the excess X-ray emission 
 was  located at 2.9$^{\prime\prime}$ South of the radio position of Sgr\,A*, 
and was coincident  with \ob{}, a moderately bright transient, discovered
 by {\sl Chandra} in July 2004 (\cite{muno05a}). 
VLA observations also detected bright radio transients symmetrically 
to \ob{} whose flux 
peaked during the Spring observing campaign (Bower et al. 2005).
The highest luminosity of \ob{}, in the 2--8\,keV energy range,  
observed by {\sl Chandra} was about 5$\times$10$^{34}$\,erg\,s$^{-1}$ (\cite{muno05b}). 
 This luminosity is too high to be explained by the X-ray emission from 
either the corona of an active star or the wind of a massive star, 
but conversely strongly favors the presence of an accreting
 compact object which could be either a white dwarf, or a neutron star, or a black hole.  
Since the typical outburst luminosity of white dwarf system is only observed up to about
 10$^{34}$\,erg\,s$^{-1}$ (e.g., \object{GK Per}, Sen \& Osborne \cite{SO98}), 
 we think that \ob{} is more likely a X-ray binary with a neutron star or
 a black hole as compact object. \\

\begin{table}[!t]
\caption{Observation log of the PN camera.}
\begin{tabular}{@{}ccc@{}c@{}}
\hline
\hline
\small{Rev.}& Start time& End time &  \small{Livetime} \\
  \#     &      (UT)           &    (UT)             &  (ks)     \\
\hline
406    & 2002-02-26T06:42:44 & 2002-02-26T17:49:55 &  ~40.0   \\ 
\hline
788    & 2004-03-28T16:46:35 & 2004-03-30T03:47:37&   105.3  \\
789    & 2004-03-30T15:39:53 & 2004-04-01T03:53:19 &  106.2  \\
866    & 2004-08-31T03:36:01 & 2004-09-01T16:03:58 &  127.5  \\
867    & 2004-09-02T03:25:38 & 2004-09-03T16:11:28 &  130.8   \\
\hline
\hline
\end{tabular}
\label{tab:log}
\end{table}

We report here X-ray eclipses from \ob{} discovered during the 
{\sl XMM-Newton} observations of Summer 2004.
 The observations of eclipses from X-ray binaries are rather rare 
with only 11 LMXB showing eclipses compared to the 80 LMXB reported 
in the catalog of Ritter \& Kolb\footnote{Version 7.4, http://www.mpa-garching.mpg.de/RKcat/}. 
 Indeed for very high inclination angles, the thickness of the accretion disc 
itself prevents us from seeing any of the X-rays at all. Eclipses are therefore 
observable only in a small number of systems where the inclination still allows us 
to see above the edge of the disk, see e.g. the review of \cite*{CS95}.
The period and duration of the eclipses can be used to constrain the mass 
of the secondary star and the mass of the compact object, respectively. \\

This paper is structured in the following manner.
In section~\ref{sec:obs} we present the {\sl XMM-Newton} observations and data
 reduction methods. Sections~\ref{sec:timing} and \ref{sec:spectral} describe
 the timing and spectral analysis of \ob, and are followed in sect. \ref{sec:eclipse}
 and \ref{sec:nature} by the study of the characteristics of the transient eclipses
 and the constraints that these impose upon the nature of the source.

\section{XMM-Newton observations and data reduction}
\label{sec:obs}

\begin{figure*}[Ht!]
\centerline{
\begin{tabular}{cc}
\includegraphics[width=5cm,angle=90]{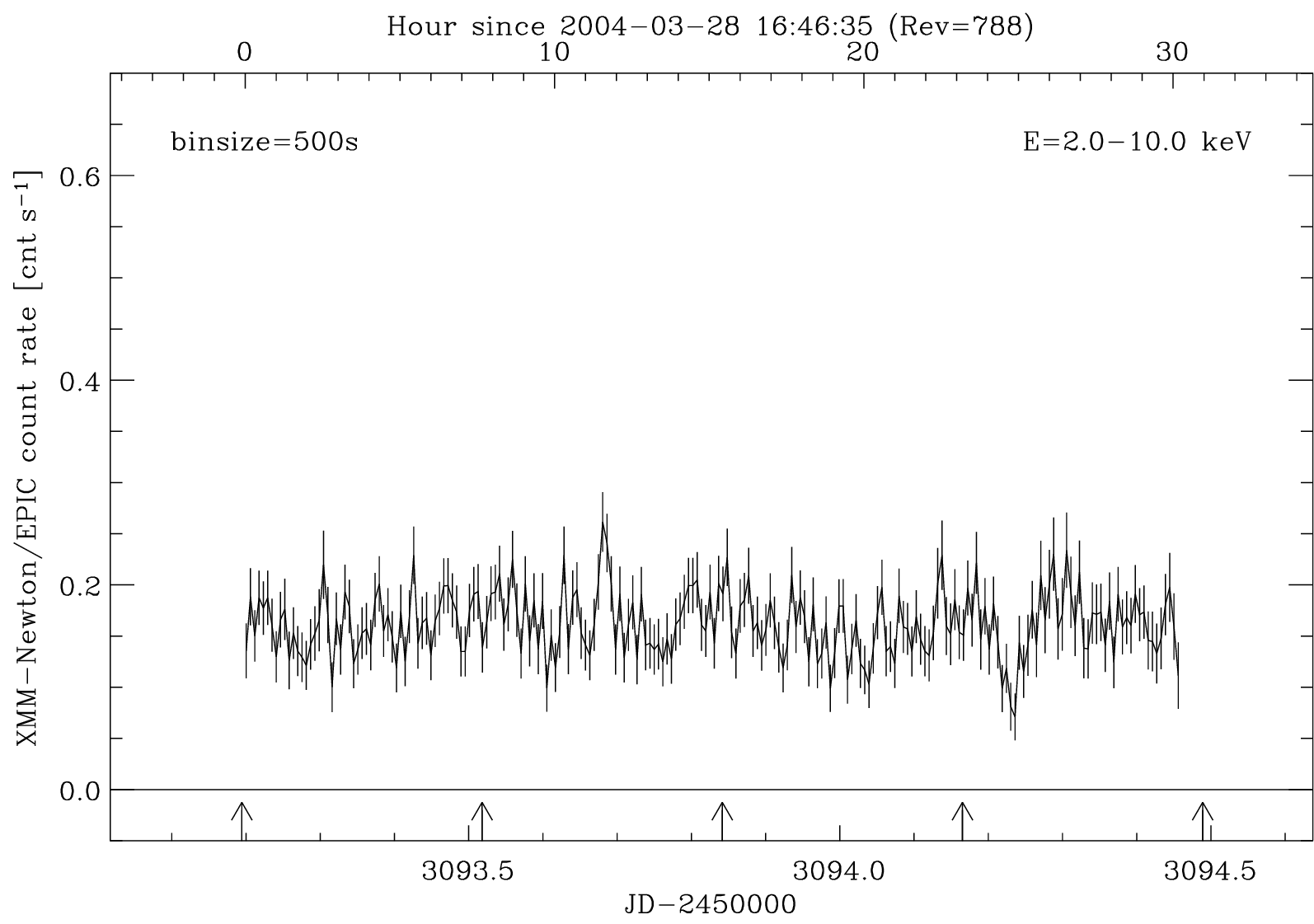} & \includegraphics[width=5cm,angle=90]{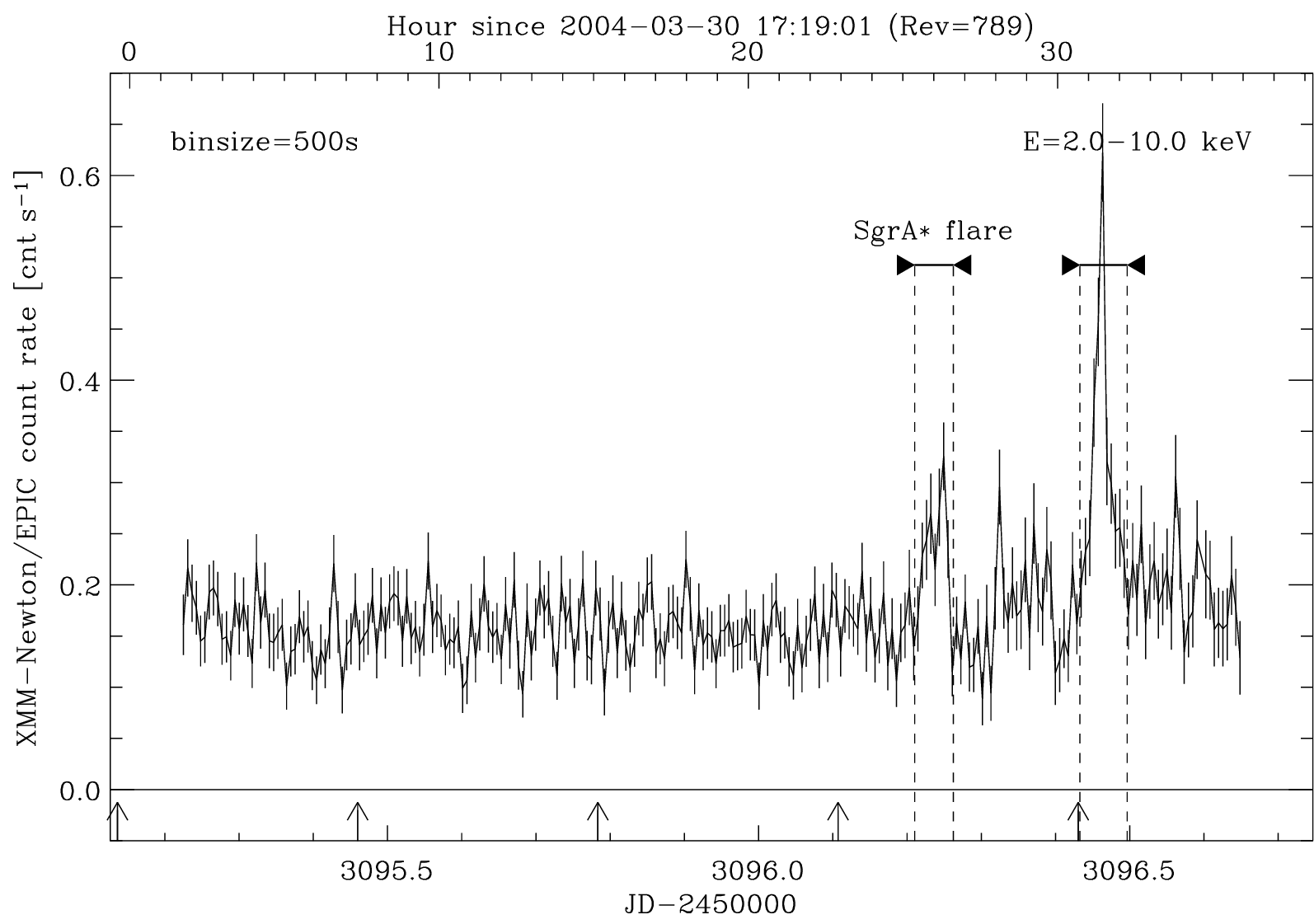}\\
\includegraphics[width=5cm,angle=90]{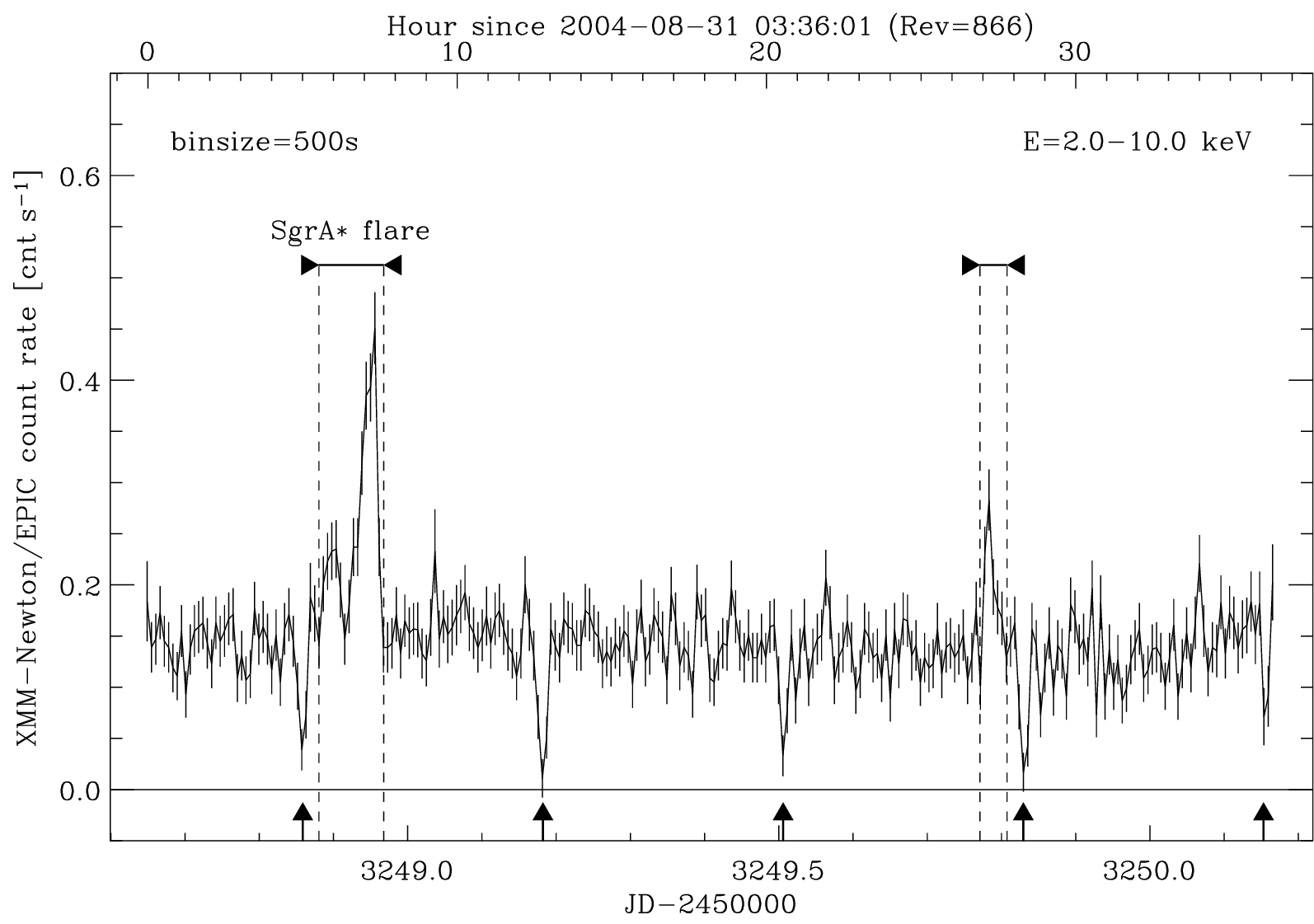} & \includegraphics[width=5cm,angle=90]{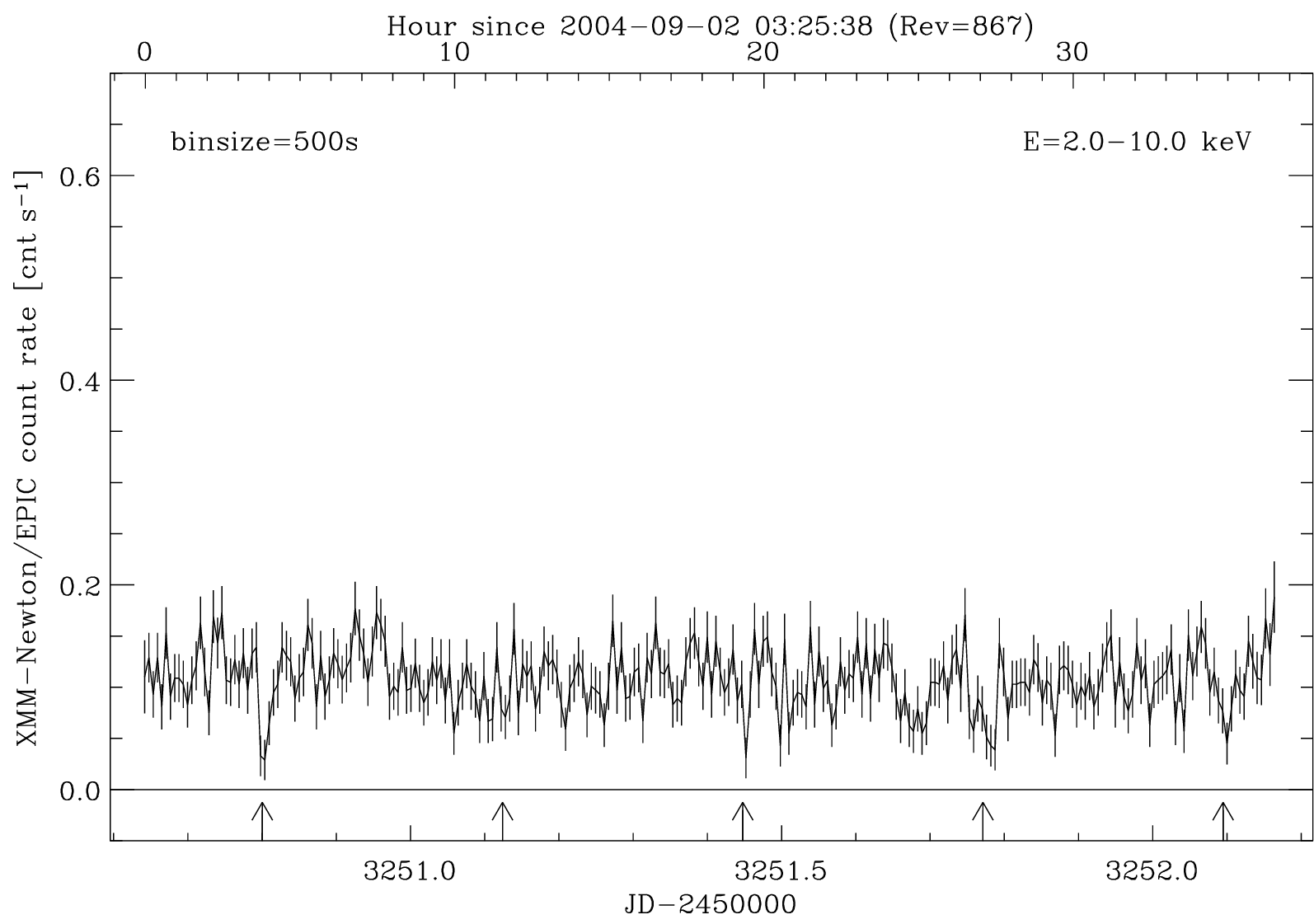}
\end{tabular}
}
\caption{EPIC light curves of \ob{} in the 2--10\,keV energy band. The quiescent X-ray emission of Sgr\,A* and the diffuse X-ray emission have been removed. The time bins are set to 500\,s, and each light curve is plot using the same scale. The horizontal arrows indicate time intervals affected by X-ray flares from Sgr\,A* (see B\'elanger et al. 2005). Vertical thick arrows point to the five X-ray eclipses observed during the revolution 866 which were used to determined the eclipse ephemeris (see $\S$\ref{sec:eclipse}). Then the predicted X-ray eclipses are marked by vertical thin arrows for the other revolutions (788, 789, and 867). While several X-ray eclipses are observed at the position of the eclipse ephemeris during revolution 867, no obvious deep eclipses are observed during the revolutions 788 and 789 (Spring 2004 observations).} 
\label{fig:lc}
\end{figure*}

The {\sl XMM-Newton} observations presented here are part of a large 
multi-wavelength campaign to monitor the flux of Sgr A*.
 The observations were made 
from March 28 to April 1 2004 (revolutions 788 and 789), 
and from August 31 to September 2 2004 (revolutions 866 and 867). 
The total exposure time is about 500\,ks.
Table~\ref{tab:log} gives the observation log of the PN camera 
with the beginning and the end time of the first and last Good Time Interval (GTI) 
of the on-axis CCD \#4, respectively. 
We also give in this table the {\sl XMM-Newton} observation of Sgr\,A* 
of February 26, 2002, where the flaring background was very low, 
where no X-ray flares were observed from Sgr\,A*, and where \ob{} was in its quiescent
 state. This latter observation will be used as reference for the X-ray flux of the 
quiescent emission of Sgr\,A* and the diffuse emission in the extraction
 region of \ob{}.
The PN camera was in Prime Full extended window 
and Prime Full window mode, during revolutions 788 and 789, 
and revolutions 406, 866, and 867, respectively.
The MOS cameras were in Prime Full window mode during all the observations. 
The data were processed with {\tt SAS} (version 6.1.0). 
We selected only X-ray events with data quality flag equal to 0. 
X-ray events with patterns 0--12 and 0--4 were used for MOS and PN, respectively.

\section{Timing analysis}
\label{sec:timing}

The position of \ob{} was determined in each {\sl XMM-Newton} observation  
using its relative angular separation from three bright X-ray point sources, namely 
\object{CXO\,J174530.0-290704}, \object{CXO\,J174607.5-285951}, 
and \object{CXO\,J174530.0-290704} (\cite{muno04}),  
 which were observed by {\sl Chandra} and {\sl XMM-Newton}.
We used the 5.1\,ks Director Discretionary Time observation of \ob{} 
made by {\sl Chandra} on 28 August 2004 (ObsID 5360, PI: F. Baganoff)
 to measure these three angular separations. 
The angular separations between \ob{} and the reference X-ray sources  
CXO\,J174530.0-290704, CXO\,J174607.5-285951, 
and CXO\,J174530.0-290704 are 79.5$\pm$0.1\,\arcsec, 
362.2$\pm$0.4\,\arcsec, and 415.3$\pm$0.3\,\arcsec, respectively.
 The positions of these three sources were obtained 
 in each {\sl XMM-Newton} observation with the SAS command {\sc edetect\_chain},
 using simultaneous detections in MOS1, MOS2, and PN images 
in the 0.3--10\,keV energy band. 
We extract for each of the five {\sl XMM-Newton} observations, 
the source events from a circular region of 10$^{\prime\prime}$-radius
 centered on the inferred {\sl XMM-Newton} position of the transient source.
 

We build the light curve of each camera from the beginning to the end 
of the first and the last GTI of the (on-axis) CCD \#4 of PN, respectively.   
We corrected the light curves for the loss of exposure due to 
the triggering of counting mode during high flaring background,  
where count rate goes beyond the detector telemetry limit. 
Using the GTI extension of the event file, we computed  
 for each time bin the ratio of the bin length 
 to exposure loss, and multiplied the count rate and corresponding 
error by this linear correction factor. 
Finally, light curves of the three detectors were added to produce 
the EPIC light curves for the four observations in the 2--10\,keV  
energy range. The contribution of the quiescent X-ray emission 
of Sgr\,A* and the diffuse X-ray emission have been removed
 by subtracting the averaged count rate observed in the same extraction 
region (0.161$\pm$0.002\,cts\,s$^{-1}$) during the February 2002 observation
 (see Table~\ref{tab:log}).
The X-ray flaring background has a negligible contribution to the total
flux in the 10$^{\prime\prime}$-radius region around SgrA* 
assuming the incident particles are distributed uniformly across the
detector.

Figure~\ref{fig:lc} shows the corresponding EPIC light curve of \ob{} in the
 2--10\,keV energy band for each of the four {\sl XMM-Newton} revolutions.
X-ray flares identified as coming from SgrA* are clearly visible and the analysis
 of these events is presented in \cite*{B05}. During revolution 866 we observed five
 periodic sharp flux troughs (indicated by thick vertical arrows in Fig.~\ref{fig:lc}). 
These features will be discussed in sect.~\ref{sec:eclipse}.

\section{Spectral analysis}
\label{sec:spectral}

\begin{figure*}[!ht]
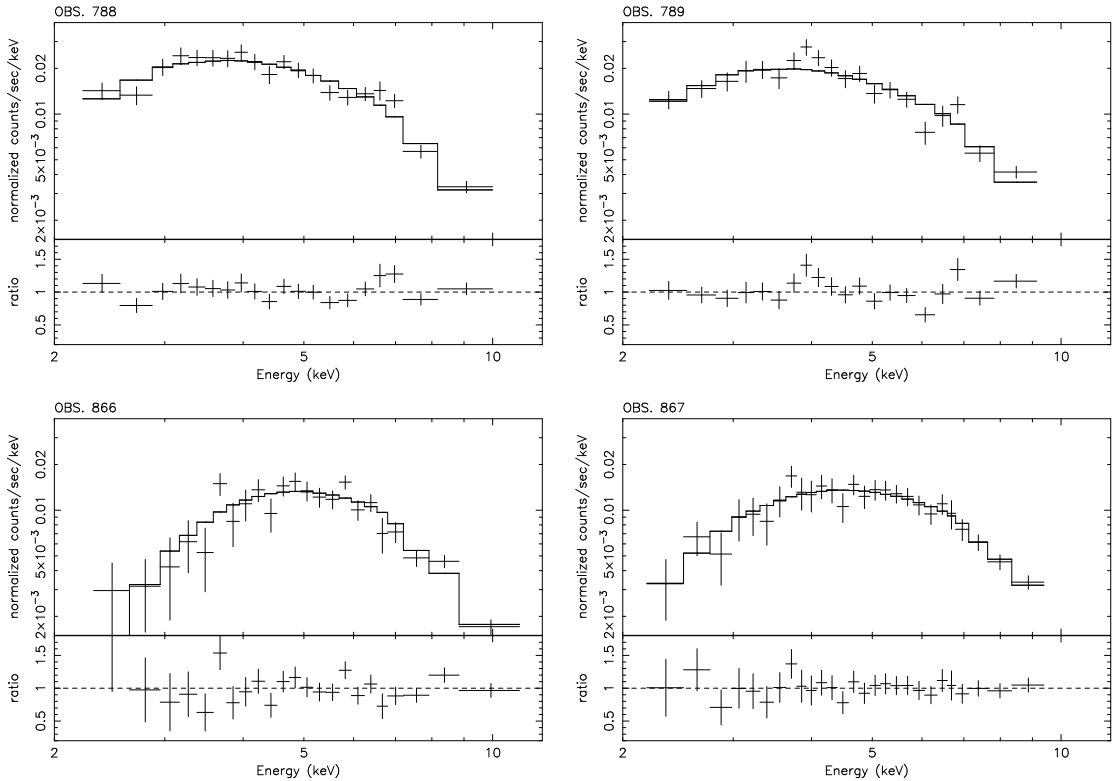

\centerline{
\begin{tabular}{cc}
\includegraphics[width=5cm,angle=-90]{3214fig2a.ps} &\includegraphics[width=5cm,angle=-90]{3214fig2b.ps}\\
\includegraphics[width=5cm,angle=-90]{3214fig2c.ps} &\includegraphics[width=5cm,angle=-90]{3214fig2d.ps}\\
\end{tabular}
}
\caption{PN spectra of \ob{}. The model is an absorbed power law index taking into account the scattering of X-rays by dust. For a direct comparison between the spectra, the energy and count rate scales are identical.} 
\label{fig:spectra}
\end{figure*}

The spectral analysis has used only PN data, 
which have a higher S/N than the MOS data, and are sensitive up to 12\,keV. 
The PN data has been cleaned from flaring background contributions  
by removing all time intervals where the PN background count rate was higher
 than 50 cts\,s$^{-1}$ and 15 cts\,s$^{-1}$ for the Spring and Summer 2004 observations, 
respectively. In addition we removed all time intervals with a significant contribution 
of X-ray flares from Sgr\,A* (marked with horizontal arrows in Fig.~\ref{fig:lc}).  
The astrophysical X-ray background (quiescent emission of Sgr\,A* and diffuse emission) 
was obtained from the {\sl XMM-Newton} observation of February 26, 2002, where both 
Sgr\,A* and \ob{} were in their quiescent state.
 With this double subtraction method, only the 
contribution of \ob{} above its quiescent state is detected. 

We use the updated X-ray absorption 
cross-sections of \cite*{Wi00} ({\sc tbabs} in {\sc xspec}).
We fit the data taking into account the scattering of X-rays by dust, 
using the {\tt scatter} model (\cite{PS95}) 
assuming a visual extinction value $A_{\rm V}$=30\,mag, as determined 
from IR observations of stars close to Sgr\,A* (e.g., \cite{Rieke89}).
The fitting parameter errors correspond to 90\% confidence ranges 
for one interesting parameter ($\Delta \chi^{2}$=2.71). 
Table~\ref{tab:fit} gives our best fit parameters for the four observations 
using an absorbed power law continuum model.
Taking into account the error bars, the photon power law indices are 
 consistent with a constant value in the four observations.
 However there is a hint of spectral hardening and decrease 
of the intrinsic luminosity during the last observation.
Interestingly, there is a significant variation of the column density 
 between Spring and Summer 2004 
(see variation of the soft emission in Figure~\ref{fig:spectra}), with the highest 
value of $N_{\rm H}$ observed during the revolution 866 of Summer 2004, 
where several deep X-ray eclipses from \ob{} are observed.

\begin{table}[!t]
\caption{Spectral fit of the PN spectrum of \ob{} for each observation 
with an absorbed power law continuum taking into account the scattering 
of X-rays by dust. 
 The observed flux ($F_{\rm X}$)
and the luminosity corrected for absorption ($L_{\rm X}$) are expressed in 
10$^{-12}$\,erg\,cm$^{-2}$\,s$^{-1}$ and 10$^{34}$\,erg\,s$^{-1}$, respectively. 
The luminosity is calculated assuming $d$=8\,kpc.}
\begin{tabular}{@{}ccccccc@{}c@{}}
\hline
\hline
\smallskip
Rev.  &   $N_{\rm H}$ &  $\Gamma$ & $\chi^{2}$/d.o.f. & $F_{\rm X}$ & $L_{\rm X}$ & \\
\#      & {\scriptsize (10$^{22}$\,cm$^{-2}$)}&  & & \multicolumn{2}{c}{2--10\,keV} \\
\hline
 788 &  6.1$\pm$1.4  & 1.98$^{+0.25}_{-0.24}$  &  22.5/17  & 1.8$^{+2.1}_{-1.2}$& 2.3$^{+1.5}_{-0.9}$ \\
 789 &  5.7$\pm$1.4  & 2.03$^{+0.26}_{-0.25}$ & 22.5/18  & 1.6$^{+1.1}_{-0.6}$&  2.0$^{+1.4}_{-0.7}$ \\
 866 &16.2 $^{+4.3}_{-3.6}$ & 2.01$^{+0.40}_{-0.38}$& 15.6/20 & 1.4$^{+1.7}_{-0.8}$& 1.9$^{+2.5}_{-1.0}$\\
 867 &9.2$^{+2.8}_{-2.3}$  & 1.60$^{+0.34}_{-0.31}$ & 10.5/24 & 1.3$^{+2.5}_{-1.2}$ & 1.8$^{+1.8}_{-0.9}$  \\
\hline
\end{tabular}
\label{tab:fit}
\end{table}

\section{X-ray eclipses}
\label{sec:eclipse}

\begin{figure*}[!ht]
\begin{center}
\begin{tabular}{cc}
\includegraphics[width=0.6\columnwidth,angle=90]{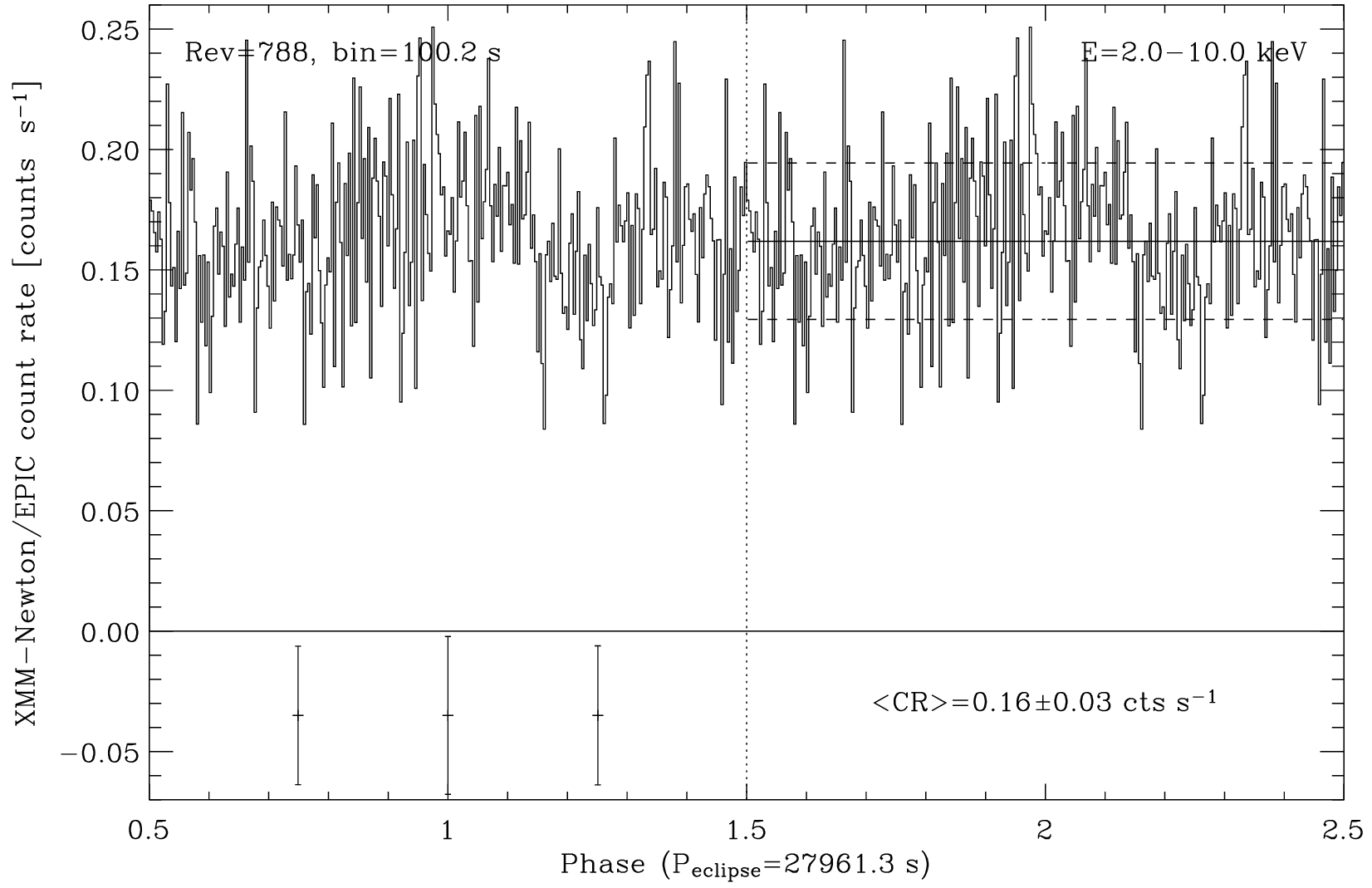} & \includegraphics[width=0.6\columnwidth,angle=90]{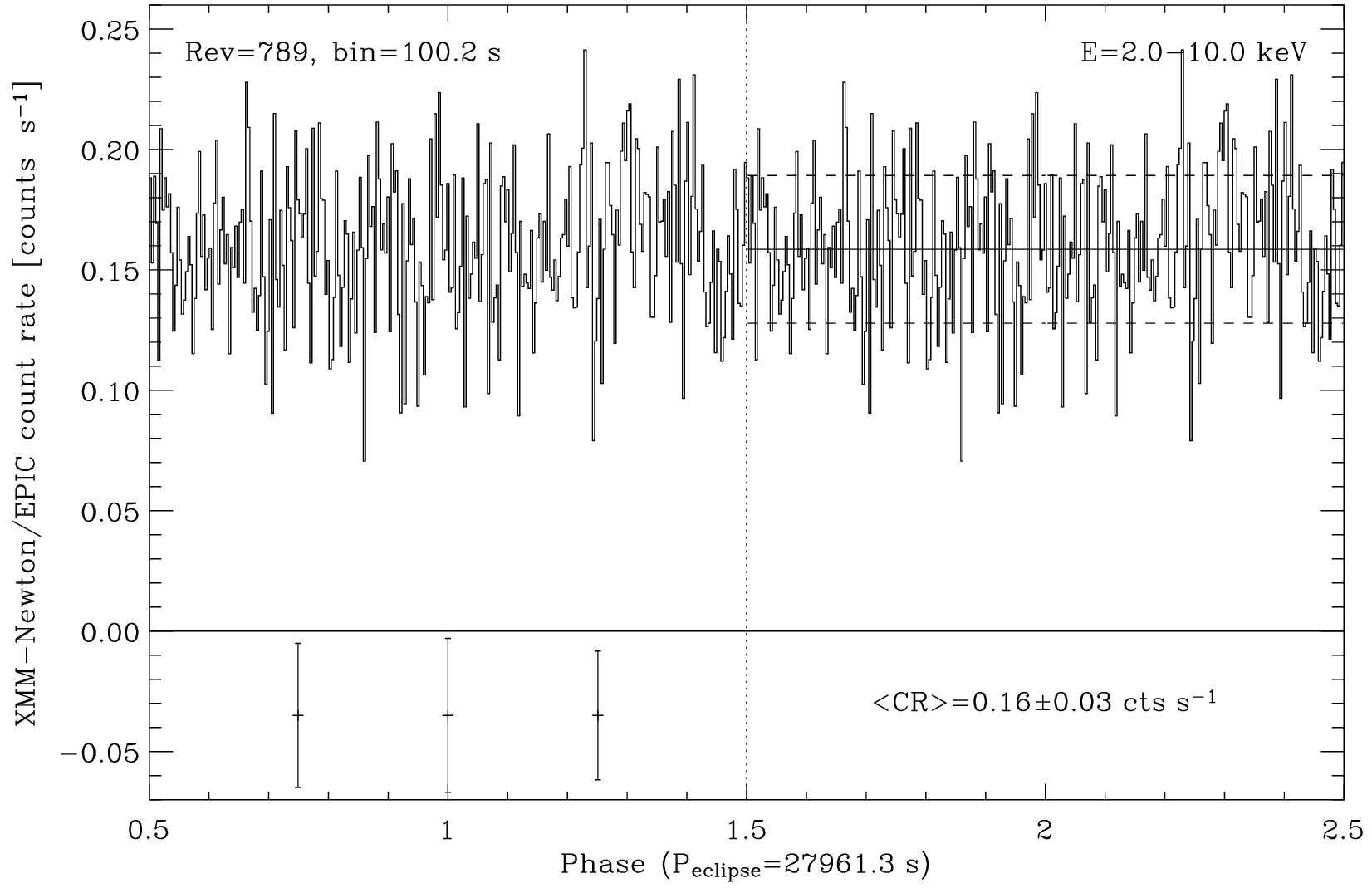} \\
\includegraphics[width=0.6\columnwidth,angle=90]{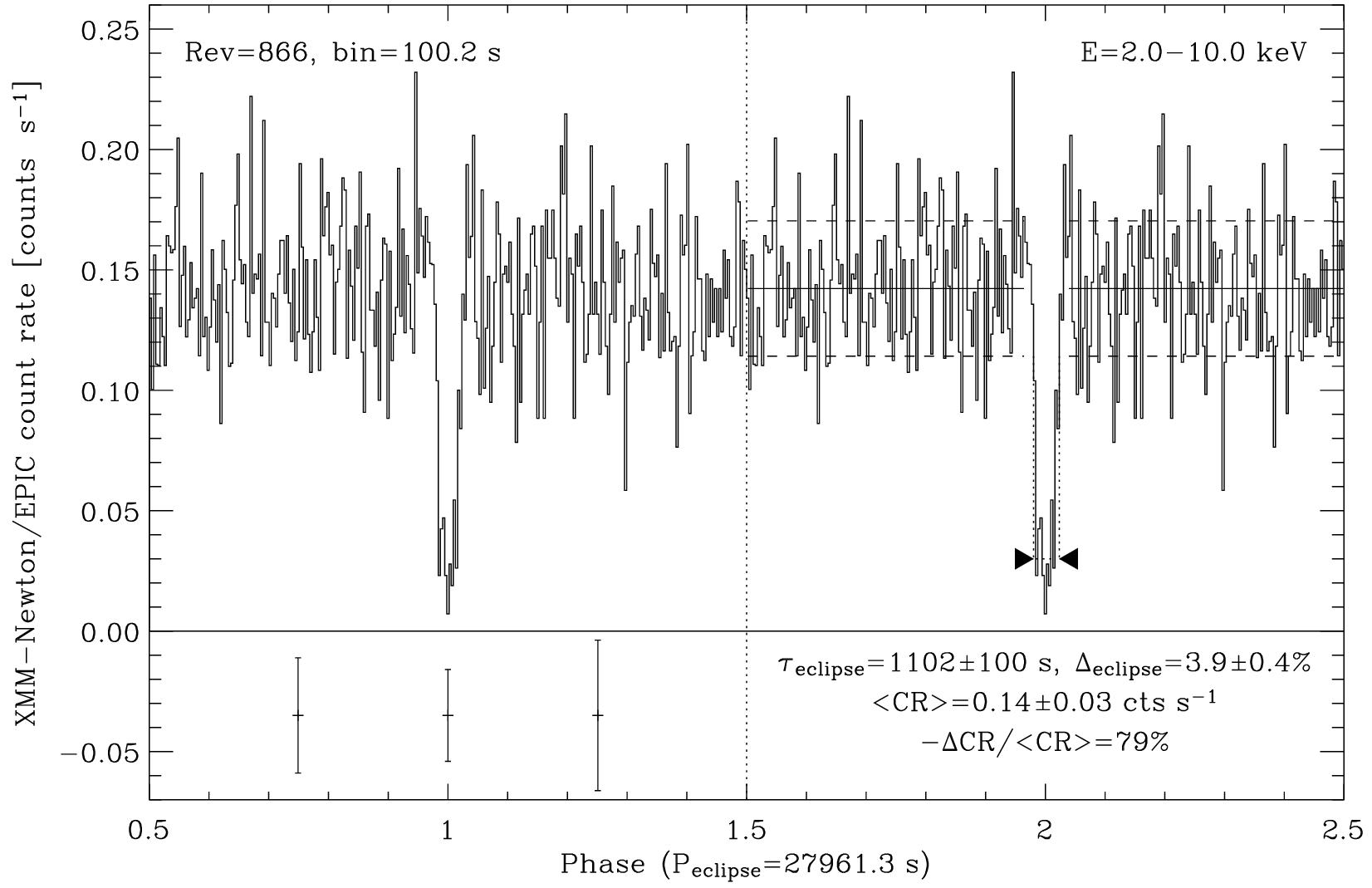} &\includegraphics[width=0.6\columnwidth,angle=90]{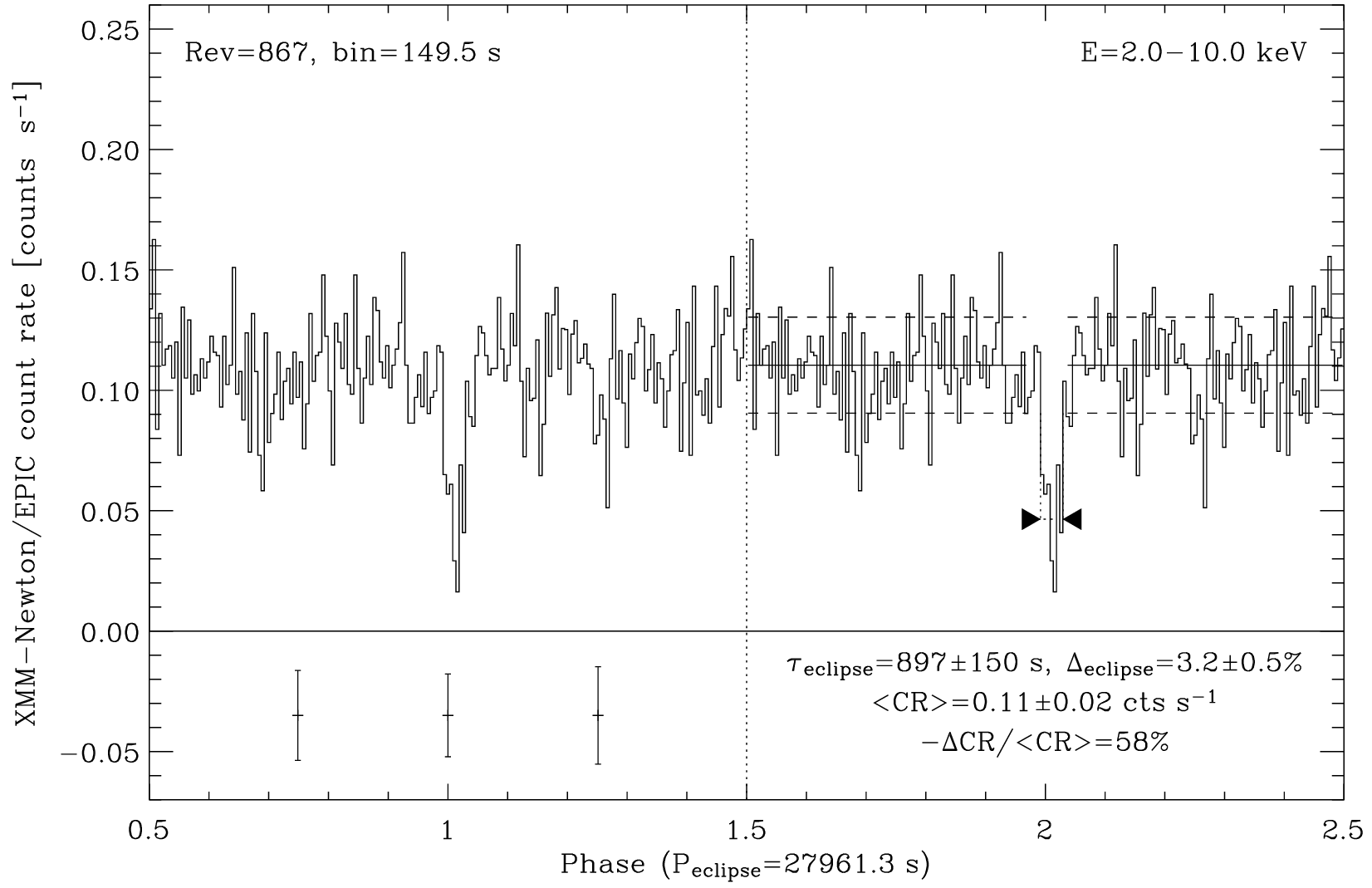}\\
\end{tabular}
\end{center}
\caption{EPIC folded light curves of \ob{}
 for revolutions 788, 789, 866 and 867 in the 2--10\,keV energy range.
Two consecutive periods are plotted.
 The horizontal continuous and dashed lines in the second period indicate the
 average count rate level, $<$CR$>$, and one sigma dispersion, respectively . 
 The duration of the eclipse $\tau_{\rm eclipse}$ is marked by arrows.  
The eclipse fraction $\Delta_{\rm eclipse}$ is defined as the ratio of the eclipse
 duration to the orbital period. The depth of the eclipse is defined by the relative 
count rate variation at the observed eclipse maximum.    
For clarity purpose, only three 1$\sigma$ error bars are indicated.}
\label{fig:folded_lc}
\end{figure*}

As reported in sect.~\ref{sec:timing}, five periodic sharp flux troughs 
are observed during revolution 866.
Since the duration of these troughs is short compared to the trough period,
we associate these features with eclipses. These eclipses are likely 
produced by the secondary star of the binary system rather than 
by optical depth effects or electron scattering in material
near the magnetic pole of the compact object (as seen in some NS or WD systems), 
which would produced broader dips. Therefore we will assume in the following 
that this eclipse periodicity is associate with the orbital motion of 
the binary system, where a compact object emits X-rays.

We determine the eclipse ephemeris using the five eclipses of revolution 866,
where the eclipses are well defined, and not affected by exposure losses
 (see vertical thick arrows in Fig.~\ref{fig:lc}). We start
from the arrival times of photons detected by MOS1+MOS2+PN, which are convolved
 with a Gaussian smoothing kernel to reduce the Poissonian noise and to produce a
 smooth shape of the light curve. Each eclipse is then fitted with a Gaussian plus
 a linear term to measure accurately the five eclipse epochs. The eclipse period
 ($P_{\rm eclipse}$) and the epoch of the first eclipse (T$_{0}$) are obtained with
 corresponding errors from mean and rms of the eclipse epoch differences. 
The best estimate of the eclipse period is obtained by minimizing its rms error by varying
the FWHM of the smoothing kernel. The minimum rms is obtained for a FWHM$\sim$760\,s,
leading to T$_{0}$=05h00m18s and $P_{\rm eclipse}=27,961\pm5$\,s.
 The Julian date epochs of the eclipses are 
$JD=(2453248.85856 \pm 0.00004) +0.323626 \pm 0.00006 \times n$,
 where $n$ is an integer. 
 The predicted eclipses are marked in Fig.~\ref{fig:lc} by thin vertical arrows. 
 While similar fainter eclipses, coinciding with the ephemeris predictions 
(see vertical thin arrows), are visible during revolution 867, 
 no obvious deep eclipses are observed during the revolutions 788 and 789
 (Spring 2004 observations) at the times predicted by the eclipse ephemeris. 
 Based on the {\it Chandra} observations performed in July and August
2004, \cite*{muno05a} proposed the possible presence of dips 
from this source with a tentative period of 7.9 h, slightly larger 
than our measurement. Our eclipse ephemeris predict well the epochs 
of the troughs observed with {\it Chandra} in July 2004.

To obtain folded light curves, the arrival times of photons detected by each instrument
 when Sgr\,A* was not flaring were folded using the eclipse period.
 Then using the GTI extension and the time intervals 
of Sgr\,A* flares, we determine the net exposure within each phase bin to calculate the count rate. 
Instrument folded light curves are then added to obtain the EPIC folded light curve. 
Figure~\ref{fig:folded_lc} shows the resulting folded light curves. 
 We compute the average count rates of the transient 
and its dispersion outside the time interval 
where the eclipse occurs. 
The average count rate level for the revolutions 788, 789, 866, and 867 
are $<$CR$>$=0.16$\pm$0.03, 0.16$\pm$0.03, 0.14$\pm$0.03, and 0.11$\pm$0.02\,cts\,s$^{-1}$,
 respectively. During each epoch, $<$CR$>$ is constant, whereas 
there is a hint of a decrease between the Spring and Summer epochs. 
Deep eclipses are visible for the revolutions 866 and 867 during Summer 2004
(Fig.~\ref{fig:folded_lc} lower panels).
The duration of the eclipse is defined as the time interval where the count rate  
is lower than one sigma from the average count rate.
Both eclipse duration measurements are consistent with about 
1,100$\pm$100\,s, which is a small fraction ($\sim$4\%) of the eclipse period. 
The eclipses are not total as suggested by the folded curves of
 the revolutions 866 and 867 (Fig.~\ref{fig:folded_lc})
where the minimum flux is not compatible with zero.
The eclipse of revolution 867 is less deep ($-\Delta CR/<CR>\sim60\%$) 
than the eclipse of revolution 866 ($\sim80\%$). 
This could suggest an extended X-ray emitting region, 
which would be only partly hidden by the secondary star. 
 As previously suggested by the light curves of revolutions 866 and 867    
there is no such deep eclipses during the Spring 2004 observations  
  (see the upper panels of Fig.~\ref{fig:folded_lc} upper panels). 
 We can exclude confidently for this epoch any 
eclipse with depth greater than 19\% 
(i.e. below one sigma dispersion of the light curve). 
 The absence of these deep X-ray eclipses in Spring 2004 
are possibly due to a displacement of the X-ray emitting region 
above the orbital plane as we will discuss in \S\ref{missing_eclipse}. 

\section{Nature of \ob}
\label{sec:nature}

We consider that the X-ray source \ob{} is a semidetached binary system,
 with a circular orbit (tidal circularization in binary system with short period),
which has a sufficiently high inclination to the line-of-sight
so as to produce X-ray eclipses of the compact object by the secondary star.
In this framework, the eclipse period ($P_{\rm eclipse}=27,961$\,s)
 is identified as the orbital period of the binary system, 
and will provide constraints on the mass of the secondary star, whereas the
 eclipse fraction ($\Delta_{\rm eclipse}=0.039$) will then constrain the mass
 of the compact object (e.g., \cite{WE04}). 

  \subsection{Upper limit on the mass of the secondary star from the eclipse period}
  \label{upper_mass}

\begin{figure}[!t]
\centering
\includegraphics[angle=90,width=\columnwidth]{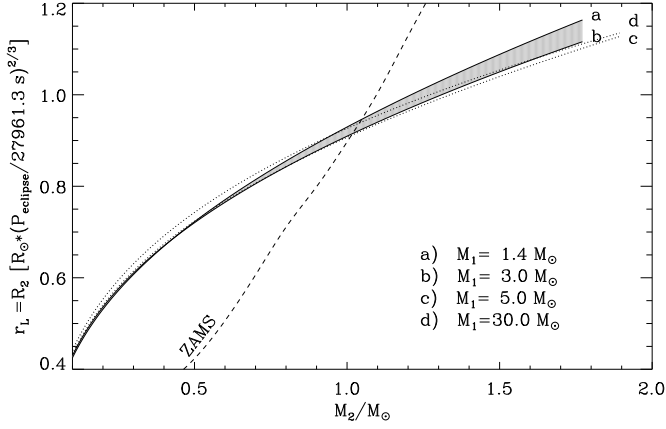}
\caption{Effective Roche lobe radius, $r_{\rm L}$, of the secondary star in a semidetached binary system with an orbital period of 27,961\,s versus the mass of the secondary star, $M_2$. Curves $a$ and $b$ (continuous lines) are for a mass of the primary of 1.4 and 3.0\,M$_\odot$, respectively (mass range of a neutron star; grey stripe); curves $c$ and $d$ (dotted lines) are for a mass of the primary of 5 and 30\,M$_\odot$, respectively (typical mass of a stellar-mass black hole). The dashed curve shows the mass-radius relation of low-mass stars in the Zero-Age Main-Sequence stage (ZAMS), with the stellar radius derived from pre-main sequence tracks with metalicity $Z=0.02$ (and $Y= 0.277$) calculated by Siess et al.\ (2000).}
\label{fig:radius_mass}
\end{figure}

We can determine a mass-radius relation of the secondary star knowing the orbital period.
The mass of the secondary star is 
\begin{equation}
\frac{M_2}{{\rm M}_\odot} = \frac{\overline{\rho}_2}{\overline{\rho}_\odot} \left( \frac{R_2}{{\rm R}_\odot} \right)^{3} {\rm ,}
\label{eq:m_2}
\end{equation}
where $\overline{\rho}_2$, $R_2$, and $\overline{\rho}_\odot$ are the mean density
 and the radius of the secondary star, and 
the solar mean density (1.41\,g\,cm$^{-3}$), respectively. 
We consider that the secondary star's radius is equal to the secondary star's
 effective Roche lobe radius\footnote{The effective Roche lobe radius is defined
 so that the sphere with this radius has the same volume that the one within the
 Roche lobe.}, $r_{\rm L}$, given by \cite*{Eg83}'s approximation:
\begin{equation}
\frac{r_{\rm L}}{a}=\frac{0.49\,q^{2/3}}{0.6\,q^{2/3}+\ln{(1+q^{1/3})}} {\rm ,}
\label{eq:r_L}
\end{equation}
where $0<q \equiv M_2/M_1< \infty$ (where $M_1$ is the mass of the primary star), 
and $a$ is the separation of the two stars, which is related to the eclipse period by 
\begin{equation}
\frac{a}{{\rm R_\odot}}=2.35 \left( \frac{P_{\rm eclipse}}{10\,{\rm h}} \right)^{2/3} \left( \frac{M_1+M_2}{{\rm M_\odot}} \right)^{1/3}  {\rm .}
\label{eq:a}
\end{equation}
Combining Eqs.~(\ref{eq:m_2})--(\ref{eq:a}) leads to: 
\begin{equation}
\frac{\overline{\rho}_2}{\overline{\rho}_\odot} = 0.66 \, \left( \frac{P_{\rm eclipse}}{10\,{\rm h}} \right)^{-2} \frac{\{0.6\,q^{2/3}+\ln{(1+q^{1/3})}\}^3}{q(1+q)} 
{\rm .}
\label{eq:rho}
\end{equation}
\noindent Applying Eq.~(\ref{eq:m_2}) in Eq.~(\ref{eq:rho}) gives the mass-radius
 relation of the secondary star knowing the orbital period:
\begin{eqnarray}
\frac{R_2}{{\rm R_\odot}} & = & 1.15 \left( \frac{P_{\rm eclipse}}{10\,{\rm h}} \right)^{2/3} \nonumber \\
                             &   & \times \frac{q^{1/3}(1+q)^{1/3}}{0.6\,q^{2/3}+\ln{(1+q^{1/3})}} \left( \frac{M_2}{{\rm M_\odot}} \right)^{1/3}  {\rm .}
\end{eqnarray}
Figure~\ref{fig:radius_mass} shows this mass-radius relation for the eclipse period,
 compared with 
the mass-radius relation of low-mass stars in the Zero-Age Main-Sequence stage (ZAMS), 
calculated by \cite*{Si00}.
 When the star evolves away from the main-sequence its stellar radius increases. 
 Assuming that the secondary star is on the ZAMS or older,
 we can see that for a large range of masses 
of the compact star (taking $M_1=1.4$--3.0\,M$_\odot$ for a neutron star,
 or $M_1=5$--$30$\,M$_\odot$ for a stellar-mass black hole),
 the mass donor star has mass $\simless\,1.05$\,M$_\odot$. 
 As shown by \cite{K01} using stellar models and allowing for different mass 
transfer rates and different system ages prior to mass transfer, the mass of a ZAMS star 
with the same spectral type is an upper limit to the donor mass.
As well, \cite{Ho01} shown that the secondary masses in CVs just above
the period gap of about 2--3\,h should be as much as $\sim$50\% lower than would be
 inferred if one assumes ZAMS donor star. 
 Therefore, the value of the mass donor star found here ($\simless\,1.05$\,M$_\odot$) 
assuming a ZAMS star is a conservative upper limit, hence, \ob{} is a {\it Low-Mass X-ray Binary} (LMXB). 

This maximum stellar mass corresponds to a mass donor star with spectral
 type later than about K0. 
For comparison a ZAMS star later than K0 (which corresponds to an age of about 150 million years) has an absolute magnitude in the infrared $K$ band 
of $M_{\rm K} > 3.50$\,mag (\cite{Si00}), which corresponds at the galactic center distance 
($d=7.94$\,kpc; \cite{E03}) where the extinction in the $K$ band is $A_{\rm K} = 2.7$\,mag (\cite{C01}),
to $K> 20.7$\,mag. The detection of the mass donor late-type star with ground adaptive optics (e.g., NACO at VLT using 
for the infrared wavefront sensor the close $K \sim 6.5$\,mag source IRS\,7 as reference star) will allow to identify 
the X-ray transient binary in the near-infrared.

 \subsection{Upper limit on the mass of the compact object from the eclipse fraction}

The eclipse fraction, i.e.\ the ratio of the eclipse duration to the orbital period,
 is geometrically related to the angular size of the path of the observer behind
 the secondary photosphere seen from the compact object by 
\begin{equation}
\Delta_{\rm eclipse}=\frac{1}{\pi} \arccos \! \left( \frac{1}{\sin i} \, \sqrt{1- \left( \frac{R_2}{a} \right)^2 } \right)  {\rm ,}
\label{eq:eclipse_fraction}
\end{equation}
where $i$ is the orbital inclination. Using Eq.~(\ref{eq:r_L}), we find that the eclipse
 fraction depends only on the orbital inclination and the mass ratio. 

\begin{figure}[!t]
\centering
\includegraphics[angle=90,width=\columnwidth]{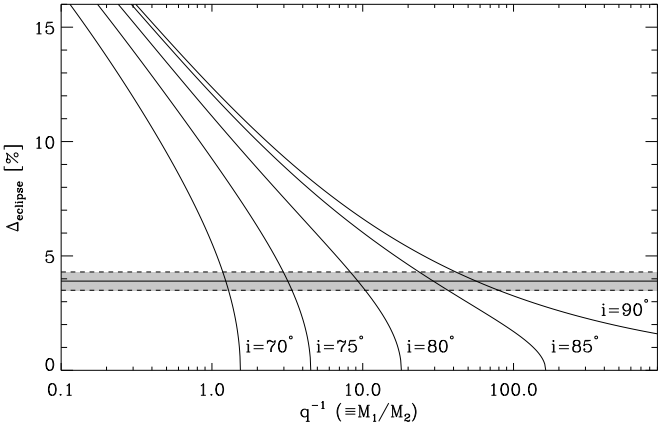}
\caption{Eclipse fraction as a function of the mass ratio for different orbital
 inclination. The horizontal grey stripe shows the observed value of the eclipse fraction.}
\label{fig:eclipse_fraction}
\end{figure}

\begin{figure}[!t]
\centering
\includegraphics[angle=90,width=\columnwidth]{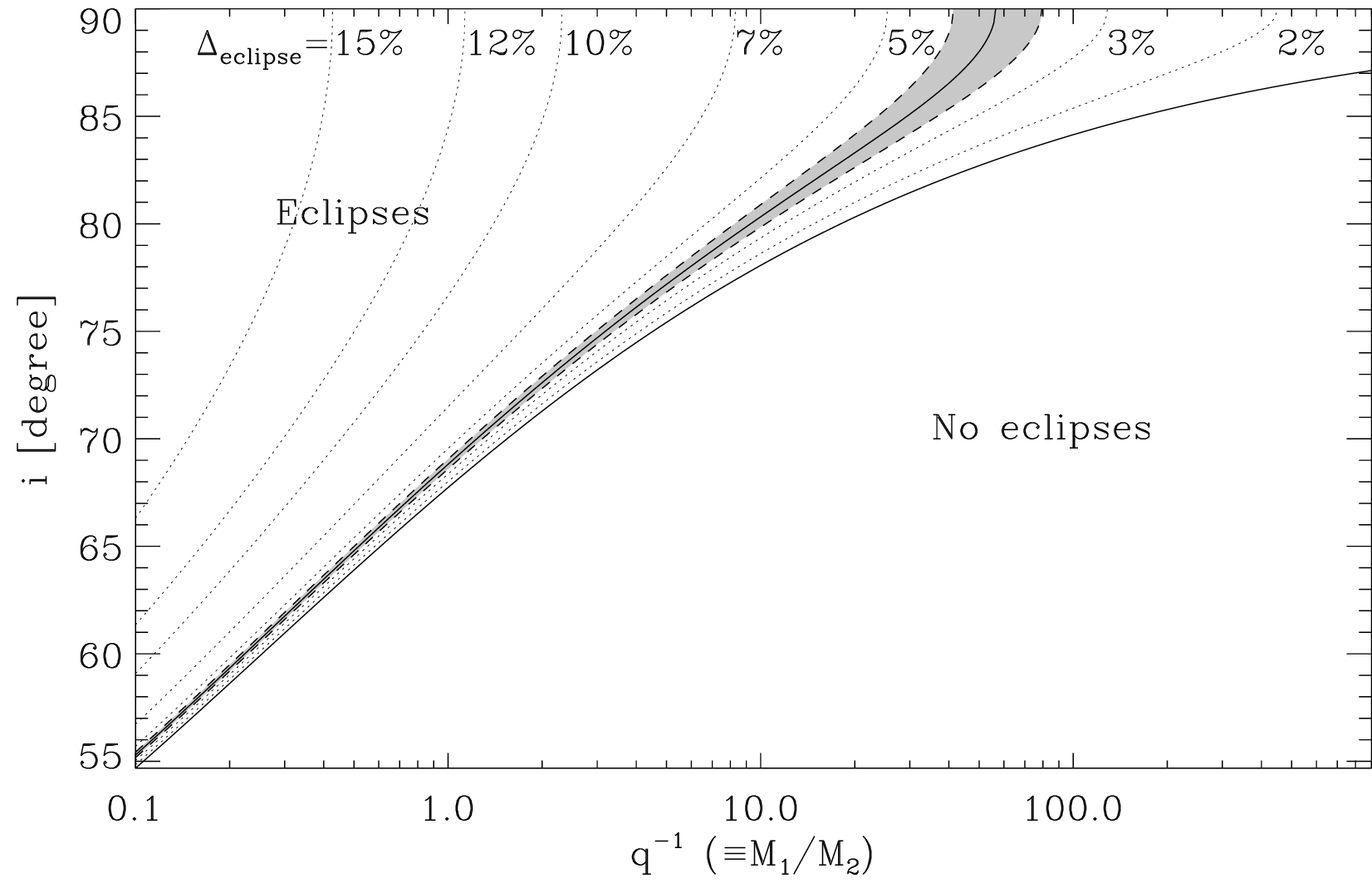}
\caption{Eclipse visibility as a function of the mass ratio and the orbital inclination. 
The continuous and dotted curves indicate the inclination values for grazing eclipses, 
and locii of constant eclipse fraction, respectively. The grey stripe shows the observed eclipse fraction.}
\label{fig:i_cr}

\centering
\includegraphics[angle=90,width=\columnwidth]{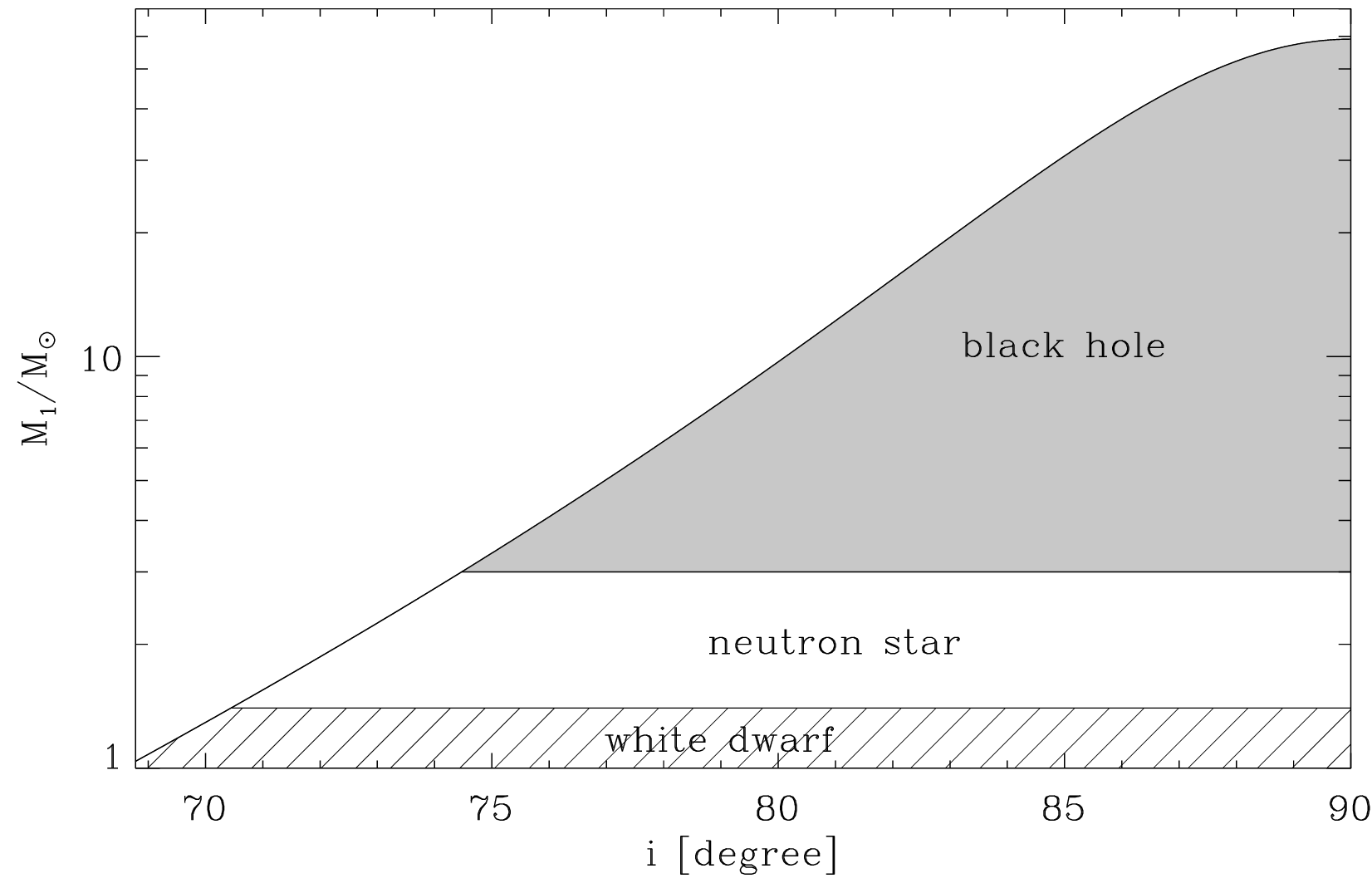}
\caption{Constrain on the mass of the compact object as a function of 
the orbital inclination when the mass of the secondary star is $< 1.05$\,M$_\odot$. 
The curve shows the mass of the compact object when the mass of the secondary star is 
equal to $1.05$\,M$_\odot$. Mass domains of white dwarfs, neutron stars, and black holes, 
are indicated.}
\label{fig:mass_inclination}
\end{figure}

Figure~\ref{fig:eclipse_fraction} shows the eclipse fraction 
as a function of $q^{-1}$ ($\equiv M_1/M_2$), for different orbital inclinations.
To each mass ratio corresponds a critical inclination where only a grazing eclipse 
is possible, below this critical inclination an eclipse of the compact source by 
the secondary star cannot occur. More generally, the proper inclination needed 
to obtain a given eclipse fraction can be derived from Eq.~(\ref{fig:eclipse_fraction}):
\begin{equation}
i= \arcsin \! \left( \frac{ \sqrt{ 1- \left( \frac{0.49\,q^{2/3}}{0.6\,q^{2/3}+\ln{(1+q^{1/3})}} \right)^2} }{\cos(\pi \Delta_{\rm eclipse})} \right) {\rm .}
\label{eq:i_cr}
\end{equation}
This equation is useful to compute the eclipse visibility as a function of the mass ratio 
and the orbital inclination (see Figure~\ref{fig:i_cr}).

The maximum eclipse duration is obtained in Fig.~\ref{fig:eclipse_fraction}
 for $i=90^\circ$, i.e.\ when the line-of-sight is parallel to the orbital plane.
 Consequently, the observed eclipse fraction provides an upper limit 
on $q^{-1}$. If the inclination is known, $q^{-1}$ is obtained by solving numerically
 the following equation:
\begin{equation}
\frac{\ln (1+q^{1/3})}{q^{2/3}} =  \frac{0.49}{\sqrt{1-\sin^2{i}\, \cos^2(\pi \Delta_{\rm eclipse})}} -0.6 {\,\rm .}
\end{equation}

The eclipse duration of 1,100\,s corresponds to $\Delta_{\rm eclipse}$=0.039,
 implying that the maximum value for $q^{-1}$, i.e.\ $M_1/M2$, is $56.5$.
 Since $M_2 \,\simless\, 1.05$\,M$_\odot$ (see \S\ref{upper_mass}), 
we obtain that the mass of the compact object is $\simless\, 59$\,M$_\odot$. 

Accretion disks around compact objects are likely not thin, and should then produce
 a permanent total X-ray eclipse when seen exactly edge-on. Therefore an orbital
 inclination of $90^\circ$ is unlikely. Figure~\ref{fig:mass_inclination} shows the
 inferred maximum mass of the compact object as a function of the orbital inclination.
 Inclinations lower than $70^\circ$ can be excluded because the luminosity of the X-ray
 transient exclude a white dwarf as primary star. We conclude that the possible values of
 the orbital inclination are between $\sim$$70^\circ$ and $90^\circ$. If the orbital
 inclination is $\sim$$70^\circ$--$\sim$$74^\circ$ a black hole can be excluded.

\begin{figure}[!t]
\centering
\includegraphics[angle=90,width=\columnwidth]{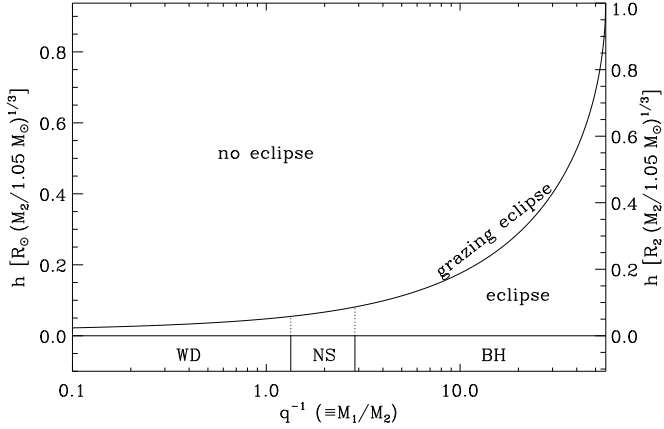}
\caption{Minimum vertical motion of the X-ray emitting region above the compact object
 needed to produce the disappearance of the eclipses when $M_2=1.05$\,M$_\odot$,
 as a function of the mass ratio. }
\label{fig:elevation}
\end{figure}

        \subsection{Constraints from the missing eclipses in Spring 2004}
        \label{missing_eclipse}

The absence of eclipse in Spring 2004 cannot be explained easily in this strict
 geometry of the binary system, where a decrease of the secondary star radius appears rather
 unlikely. We investigate a vertical motion of the X-ray emitting region from the orbital plane
 to a region located above the compact object. 
In such a configuration, the critical inclination, $i_{\rm cr}$, which produces grazing eclipses, 
is related to the vertical elevation, $h$, of the X-ray emitting region above the orbital plane by
\begin{equation}
h= \frac{R_2-a \sqrt{1-\sin^2 i_{\rm cr}}}{\sin i_{\rm cr}}  {\, \rm .}
\label{eq:z}
\end{equation}
Taking $i_{\rm cr}$ equal to the orbital inclination constrained from Eq.~(\ref{eq:i_cr})
 provides the value of the minimum vertical elevation needed to suppress the eclipses.
 Replacing in Eq.~(\ref{eq:z}), $a$ and $\sin i$ by Eq.~(\ref{eq:a}) 
and (\ref{eq:i_cr}), respectively, leads to
\begin{eqnarray}
\frac{h}{{\rm R_\odot}} & = & 2.35 \left( \frac{P_{\rm eclipse}}{10\,{\rm h}} \right)^{2/3} \left( \frac{1+q}{q} \right)^{1/3} \nonumber \\
 & & \times \left( \frac{\cos(\pi\Delta_{\rm eclipse})}{\sqrt{(\frac{R_2}{a})^{-2}-1}} - \sqrt{\frac{(\frac{R_2}{a})^2-\sin^2(\pi\Delta_{\rm eclipse})}{1-(\frac{R_2}{a})^2}} \right) \nonumber \\
 & & \times \left( \frac{M_2}{{\rm M_\odot}} \right)^{1/3}  {\rm ,}
\label{eq:h}
\end{eqnarray}
where $R_2/a$ is a function of the mass ratio only, see Eq.~(\ref{eq:r_L}). 
Using $M_2 \,\simless\, 1.05$\,M$_\odot$ provides an upper limit on
 the minimum value of $h$ needed 
to produce the disappearance of the eclipses. 

Figure~\ref{fig:elevation} shows $h$ as a function of the mass ratio for
 $M_2=1.05$\,M$_\odot$. If the compact object is a neutron star a vertical elevation
 of the X-ray emitting region of only $h\sim0.1$\,R$_\odot\sim0.1$\,R$_2$ is sufficient
 to suppress the eclipses by the secondary star. If the compact object is a black hole
 a greater vertical elevation is needed.

\section{Discussion}\label{sec:discussion}

From the {\sl Chandra} observations (\cite{muno05a}), the minimum 2--8\,keV luminosity
of \ob{} at 8\,kpc was determined to be less than 2 $\times$ 10$^{31}$\,erg\,s$^{-1}$
 (\cite{muno05a}), and the maximum to be about 5 $\times$ 10$^{34}$\,erg\,s$^{-1}$
 (\cite{muno05b}). This large amplitude variation corresponds to a factor of at least
 2,500. We infer a minimum 0.5--10 keV luminosity of less
than 3.5 $\times$ 10$^{31}$\,erg\,s$^{-1}$. Such a low luminosity lends support to the
interpretation that the compact object of this binary system is more
likely to be a black hole than a neutron star.
 Indeed, the quiescent 0.5--10 keV X-ray luminosities of neutron stars  
 are about 10$^{32}$--10$^{34}$\,erg\,s$^{-1}$,
 i.e., 100 times higher than those for black
holes (10$^{30}$--10$^{33}$\,erg\,s$^{-1}$),
 as shown by \cite*{Ga01}. 
In addition, {\sl VLA} observations have revealed a radio outburst coincident with
 the X-ray position of this source, with a peak intensity of 100\,mJy at 7\,mm (45 GHz) 
 (\cite{Bo05}). The peak radio emission occurred in 
March 2004 and continued decaying in its surface brightness until the start 
of the September 2004 campaign (\cite{Bo05}).
Therefore, the minimum luminosity and the radio outburst 
strongly suggest  that this source is more likely to
 contain a black hole than a neutron star primary 
(e.g., \cite{FK01}). 

We observed with {\sl XMM-Newton} X-ray eclipses from \ob{} with a period of
 27,961 $\pm$ 5\,s (about 7.76 hours) in the period from August 31 to September 2 2004
 but did not however detect such a feature in the earlier observations of March
28 to April 1 2004. In the framework of eclipsing semidetached binary systems,
 we show that the 27,961$\pm$5\,s eclipse period constrains the mass of the assumed
 main-sequence secondary star to less than 1.0\,M$_\odot$.
 Therefore, we deduce that \ob{} is a {\sl low-mass X-ray binary}.
 Moreover the eclipse duration (1,100$\pm$100\,s)
 constrains the mass of the compact object to less than about
 60\,M$_\odot$, 
which is consistent with a stellar mass black hole or a neutron star. 

 The obtained  {\sl XMM-Newton} spectra are well described with an absorbed  
power law continuum. The column density values on the line-of-sight and the
2--10\,keV luminosities confirm that this source is an X-ray binary, 
with a black hole or a neutron star as compact object, located at or 
near the Galactic Center. The 2--10\,keV luminosity is almost constant with
 1.8--2.3 $\times$ 10$^{34}$\,erg\,s$^{-1}$ over the four observations,
 assuming a distance of 8\,kpc. While the power law indices ($\Gamma$=1.6--2.0) 
are consistent within the statistical uncertainty over the four observations,
 we observed a significant increase of the column density during the Summer 2004
 observations, i.e. when X-ray eclipses are observed.

We propose to explain the absence and the presence of deep X-ray eclipses 
 during the Spring 2004 and Summer 2004 observations, respectively,
 without any significant change of the X-ray luminosity (see Table~\ref{fig:spectra}) 
by a different position of the X-ray emitting region.  
In Spring 2004 observations the absence of X-ray eclipses could be due to
 the vertical movement of the centroid
            of the X-ray emission from a point on the orbital plane
            to a position above the compact object  (the offset being
            at least 0.1 times the radius of the secondary star for a
            neutron star and more for a black hole primary).
In such a configuration, our line-of-sight to the X-ray emitting region
does not intercept the upper layer of
 the accretion disk, which is consistent with the lower column density value 
observed in the Spring 2004 observations.
 We propose that this displacement
of the X-ray emitting region could be linked to the presence of a jet.
When the jet is launched, a part of the hot plasma  
accreting towards the central compact object is lift off above the accretion disk, 
 extending consequently the X-ray emission region well above the limb 
of the secondary star. On the contrary in Summer 2004, the X-ray emission 
from the base of the jet becomes almost negligible and most of the X-ray emission comes 
from the mid-plane accretion disk which is eclipsed by the secondary star. 
This scenario is compatible with the significant increase in the column density
 during the Summer 2004 observations, since here the line-of-sight to the X-ray
 emitting region intercepts the upper layer of the accretion disk.

Radio observations of this transient object strongly support 
the appearance of a jet. 
Indeed, two new transient radio sources 
 were discovered by the VLA at 7\,mm on March 2004 
symmetrically to the position of the X-ray source 
(\cite{Bo05}, see also fig.~1 in \cite{muno05a}).
The peak of the radio emission occurred in 
March 2004 and continued decaying in its surface brightness
 until the start of the September 2004 campaign (\cite{Bo05}). 
 Also VLA observations at 6\,cm show clearly 
the brightening of this transient radio source in the 2004 epoch  
 compared to the 1983 epoch (Yusef-Zadeh, Cotton \& Bower 2005, in preparation). 
These two bright radio sources
 (as well as  X-ray reflection off the orbiting gas reported by \cite{muno05b}) 
most likely associated with the binary system \ob{} 
can be the imprint of radio emission from a collimated bipolar outflow, as observed 
in microquasars (e.g., \cite{spencer97}),
where a mass ejection is triggered by an instability in the accretion 
disk of a binary system, with the primary star being a black hole as suggested above. 
In conclusion, \ob{} appears to be a LMXB with more likely 
a black hole as compact object, which lead to an inclination
angle of the system greater than 75$^{\circ}$ according to 
our study. Sensitive follow up observations of this embedded X-ray binary 
with ground based adaptive optic system 
should be able to detect the mass donor star in the near-IR 
and allow us to impose more restrictive constraints on the
parameters of this transient eclipsing binary system.

\begin{acknowledgements}
This work is based on observations obtained with {\sl XMM-Newton}, 
an ESA science mission with instruments and contributions directly 
funded by ESA Member States and the USA (NASA). 
The authors would like to thank the referee for useful comments.  
 D.P. is supported by a MPE fellowship.
\end{acknowledgements}


\end{document}